\documentclass[prl,aps,twocolumn,superscriptaddress]{revtex4}

\usepackage{amsbsy,amssymb,amsmath,bm}
\usepackage{epsfig,rotate}
\usepackage{fancyhdr}

\usepackage{graphicx,color}% Include figure files
\usepackage{dcolumn}% Align table columns on decimal point
\usepackage{bm}% bold math
\usepackage{amssymb}

\begin{document}

%\preprint{Preprint}

\title{Electronic transport in the Coulomb phase of the pyrochlore spin ice}

\author{Gia-Wei Chern}
\affiliation{Theoretical Division, T-4 and CNLS, Los Alamos National Laboratory, Los Alamos, NM, 87545, USA}
\affiliation{Department of Physics, University of Wisconsin, Madison, Wisconsin 53706, USA}

\author{Saurabh Maiti}
\affiliation{Department of Physics, University of Wisconsin, Madison, Wisconsin 53706, USA}

\author{Rafael M. Fernandes}
\affiliation{School of Physics and Astronomy, University of Minnesota, Minneapolis, MN 55455, USA}

\author{Peter W\"olfle}
\affiliation{Institute for Condensed Matter Theory and Institute for Nanotechnology,
Karlsruhe Institute of Technology, D-76128 Karlsruhe, Germany}

\date{\today}

\begin{abstract}
We investigate the transport properties of itinerant electrons interacting with 
a background of localized spins in a correlated paramagnetic phase of the pyrochlore lattice.
We find a residual resistivity at zero temperature due to the scattering of electrons by the static dipolar 
spin-spin correlation that characterizes the metallic Coulomb phase. As temperature increases, thermally excited
topological defects, also known as magnetic monopoles, reduce the spin correlation, hence suppressing electron scattering.
Combined with the usual scattering processes in metals at higher temperatures, this mechanism yields a non-monotonic resistivity, 
displaying a minimum at temperature scales associated with the magnetic monopole excitation energy. 
Our calculations agree quantitatively with resistivity measurements in Nd$_{2}$Ir$_{2}$O$_{7}$, 
shedding light on the origin of the resistivity minimum observed in metallic spin-ice compounds. 
\end{abstract}

\maketitle

The interaction between itinerant electrons and localized moments continues to attract considerable interest
as model systems to understand non-Fermi liquid behaviors and unusual transport phenomena.
Since the pioneering work of Kondo on the resistivity minimum in metals~\cite{kondo}, the Kondo-lattice model
has become the paradigm for a new class of problems~\cite{doniach}. This simple Hamiltonian describing the competition between 
Kondo screening of individual moments and direct exchange among localized spins plays an important role in the
physics of heavy-fermion materials~\cite{hewson}. Novel magneto-transport phenomena such as colossal magnetoresistance~\cite{jin94}
and anomalous Hall effect~\cite{nagaosa10} emerge in a closely related double-exchange model in which the Kondo-screening
is replaced by the effect of a ferromagnetic Hund's rule coupling~\cite{zener}.

\begin{figure}[t]
\includegraphics[width=0.95\columnwidth]{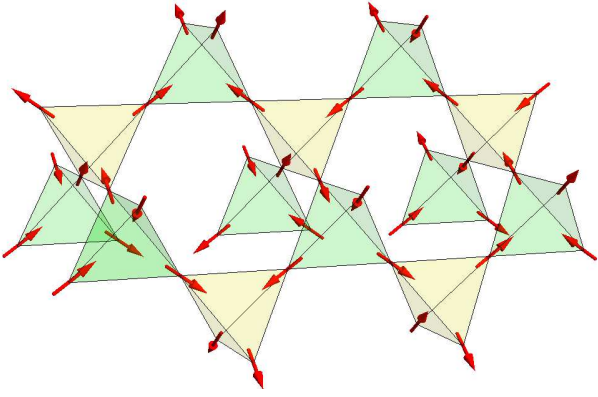}
\caption{(Color online) A fragment of the pyrochlore lattice. In the Coulomb phase, local spin arrangement
on each individual tetrahedra obeys the so-called ice rules with two spins pointing in and two spins pointing out.
\label{fig:spinice}}
\end{figure}

Adding geometrical frustration to Kondo-lattice models opens a new avenue for exploring unconventional phase transitions and many-body states.
For example, a spin-liquid phase with broken time-reversal symmetry and spontaneous Hall effect
was recently observed in the pyrochlore iridate Pr$_2$Ir$_2$O$_7$~\cite{machida10}. The conduction $5d$ electrons in this compound
coexist with the localized $4f$ moments at the Pr sites that form a frustrated pyrochlore lattice (Fig.~\ref{fig:spinice}).
A strong easy-axis anisotropy forces the moments to point along the local $\langle 111 \rangle$ directions.
When spin interactions are restricted to nearest-neighbor ferromagnetic exchange $J_F$,
an extensively large number of Ising configurations satisfying the ``ice rules'' are degenerate ground states~\cite{spinice}.
As temperature tends to zero, this degeneracy is eventually lifted by residual perturbations, 
such as spin-electron coupling~\cite{ikeda08}.  However, spatial spin correlations exhibit an unusual power-law decay over a wide
temperature range below the Curie-Weiss constant $\Theta_{\rm CW} \sim J_F/k_B$. 
Elementary excitations in this so-called Coulomb phase~\cite{henley10} are defect tetrahedra 
that violate the ice rules and carry a finite magnetic charge~\cite{castelnovo08,ryzhkin05}.

Recently, an unexpected resistivity minimum, similar to the one seen in Kondo systems, 
was observed in the paramagnetic phase of the metallic pyrochlore iridates Pr$_2$Ir$_2$O$_7$ and Nd$_2$Ir$_2$O$_7$~\cite{nakatsuji06,sakata11}.
Although quantum fluctuations of these Ising-like moments were shown to be non-negligible~\cite{onoda10},
the large easy-axis anisotropy suppresses the Kondo screening. On the other hand, a non-Kondo mechanism
for the resistivity minimum was suggested by a recent numerical study of a strong-coupling Kondo-lattice
model assuming classical Ising spins on the pyrochlore lattice~\cite{udagawa12}.
In particular, their cellular dynamical mean-field calculation suggests a clear
correlation between the position of resistivity minimum and the onset temperature of ice rules.
However, up to date, there is not a transparent theory relating the resistivity of the Coulomb phase to its unique magnetic correlations
and excitations.

In this paper, we present an analytical theory for electron transport in the Coulomb phase of spin ice. 
In the continuum approximation, the problem is recast into that of a correlated random ``magnetic'' field locally coupled to 
a Fermi sea of electrons. The resulting electronic scattering is the dominant transport mechanism at low temperatures. 
We show that thermally excited monopoles reduce the spin-spin correlation and suppress the electron scattering rate. 
Combined with the other scattering processes taking place in the metallic compounds - due, for instance, to electron-phonon
interaction - we obtain a resistivity minimum at temperature scales associated with the excitation 
energy of the magnetic monopoles. We compare our calculations with the resistivity data on Nd$_{2}$Ir$_{2}$O$_{7}$
and find a good quantitative agreement.  Our results show that strong geometrical frustration combined with topological excitations
give rise to a resistivity minimum of non-Kondo character in metallic spin ice.

We start with a ferromagnetic Kondo-lattice model on the pyrochlore lattice:
\begin{eqnarray}
    \mathcal{H} &=& -t\sum_{\langle ij\rangle, \alpha}\left(c^{\dagger}_{i,\alpha} c^{\phantom{\dagger}}_{j,\alpha} + \mbox{h.c.}\right)
        \nonumber \\
    & &  - J_H \sum_{i,\alpha\beta} \mathbf S_i\cdot\bm\sigma_{\alpha\beta}\, c^{\dagger}_{i, \alpha} c^{\phantom{\dagger}}_{i,\beta}
    - J_F \sum_{\langle ij \rangle} \mathbf S_i\cdot\mathbf S_j.
\end{eqnarray}
Here the first term describes electron hopping between nearest-neighbor sites, $t$ is the hopping integral, 
and $c^{\dagger}_{i,\alpha}$ creates an electron with spin $\alpha = \uparrow,\downarrow$ on site $i$. The itinerant electrons interact
with the localized spins $\mathbf S_i$ through an on-site Hund's coupling $J_H$; here $\bm\sigma_{\alpha\beta}$ is a vector of
Pauli matrices. As discussed above, the magnetic moments $\mathbf S_i = \tau_i S\,\hat{\mathbf e}_i$ are forced to point along the
local $\hat{\mathbf{e}}_{i}=\langle 111 \rangle$ axes by a strong easy-axis anisotropy; $\tau_i = \pm 1$ is an Ising variable. 
The last term represents a ferromagnetic ($J_F>0$) exchange interaction between neighboring moments.

We first consider the ground state of the localized spin system. The exchange energy of a single tetrahedron is minimized by six 
different Ising states with two spins pointing in and two pointing out of the tetrahedron. This 2-in-2-out rule is analogous to the 
Bernal-Fowler rule for water ice~\cite{spinice}. As explained by Pauling, the very large number of configurations satisfying the ice rule 
in a macroscopic sample gives rise to a measurable residual entropy at low temperatures~\cite{pauling}. Because of this huge degeneracy, 
spins remain disordered even at temperatures well below the exchange energy scale $J_F$.

Remarkably, spatial correlation of spins in this disordered yet highly constrained phase exhibits a power-law decay. This is because the
local 2-in-2-out ice rule translates into a divergence-free condition $\nabla\cdot\mathbf B = 0$ for the coarse-grained
magnetization field $\mathbf B(\mathbf r) = \sum_{i\in\mathcal{V_{\mathbf r}}}\mathbf S_i/\Omega$,
where $\mathcal{V}_{\mathbf r}$ denotes a coarse-graining block of volume $\Omega \sim p_F^{-3}$ that contains several tetrahedra.
In momentum space this constraint becomes $\mathbf k\cdot\mathbf B(\mathbf k) = 0$, indicating that only transverse fluctuations are allowed.
Consequently, the static spin correlator has a dipolar form $\langle B_\mu(\mathbf r) B_\nu(0)\rangle \sim (\delta_{\mu\nu} -
3\hat r_\mu \hat r_\nu)/r^3$ for large~$r$.

The low-energy excitations of this paramagnetic phase are topological defects that carry magnetic charges, namely, tetrahedra 
with 3-in-1-out or 1-in-3-out spin configurations. Since these emergent magnetic monopoles are sources and sinks
for the magnetization field, hoppings of thermally excited monopoles modify the spin correlator at finite temperatures.  
In the low-$T$ hydrodynamic regime, the relaxation of the magnetization by monopole motion
is governed by a stochastic differential equation~\cite{ryzhkin05,castelnovo12}
\begin{eqnarray}
    \label{eq:db_dt}
    \frac{\partial \mathbf B}{\partial t} = -2\mu n_m \Phi \mathbf B + D \bm\nabla n_m + \bm\zeta(t),
\end{eqnarray}
Here $n_m = \nabla\cdot\mathbf B$ is the monopole density, $\mu$ and $D$ are the monopole mobility and diffusion constant,
respectively, $\bm\zeta(t)$ is a Gaussian noise source, $\Phi = (16/\sqrt{3})\, a T$ indicates the entropic origin of monopole
drift~\cite{ryzhkin05}, and $a$ is the lattice constant. The first term in the right-hand side of Eq.~(\ref{eq:db_dt}) implies a
magnetic relaxation time $\tau_m \sim 1/\mu n_m \Phi$. Since the mobility $\mu \sim 1/T$ at low temperatures~\cite{castelnovo12},
the combination $\mu\Phi$ is a relatively weak function of $T$, and the relaxation time is governed by the monopole density:
$n_m\sim e^{-\Delta/T}$ and diverges exponentially $\tau_m(T) = \tau_{m,0} \,e^{\Delta/T}$~\cite{jaubert09}; here $\Delta$ is the
activation energy for creating monopoles. Eq.~(\ref{eq:db_dt}) gives rise to a time-dependent correlation function~\cite{henley10,conlon09}:
\begin{eqnarray}
    \label{eq:B-corr}
    & &\!\!\langle B_\mu(\mathbf k,t) B_\nu(-\mathbf k, 0) \rangle = \frac{1}{K}
    \left(\delta_{\mu\nu} - \frac{k_\mu k_\nu}{|\mathbf k|^2}\right) e^{-|t|/\tau_m} \nonumber \\
    & & \qquad\quad + \frac{1}{K}\frac{k_\mu k_\nu}{|\mathbf k|^2} \frac{\kappa^2}{(|\mathbf k|^2 + \kappa^2)}
    e^{-(1/\tau_m + D |\mathbf k|^2)|t|},
\end{eqnarray}
where $K$ is the stiffness constant of the flux field, $\kappa^{-1} = \ell_D \propto \sqrt{T/n_m}$ is the Debye screening
length of monopoles. The two terms in Eq.~(\ref{eq:B-corr}) denote the transverse and longitudinal
components, respectively.

\begin{figure}[b]
\includegraphics[width=0.76\columnwidth]{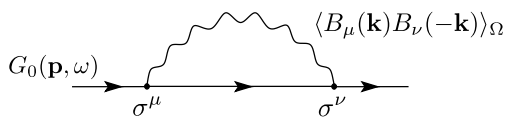}
\caption{Lowest-order diagram of the electron self-energy $\Sigma_{\alpha\beta}(\mathbf p, \omega)$.
The solid and wavy lines denote the electron propagator $G_0$ and the spin correlator~(\ref{eq:B-corr}), respectively.
\label{fig:self-energy}}
\end{figure}

Since the transport properties are dominated by conduction
electrons near the Fermi surface, their wavefunction has a
characteristic wave length $p_F^{-1}$ which is much larger than the underlying lattice constant. 
Considering the low electron density of the iridate compounds, we can choose a coarse-graining block $\mathcal{V}$
whose volume $\Omega \sim p_F^{-3}$.
We thus consider a model Hamiltonian describing electrons interacting with the averaged
magnetization field:
\begin{eqnarray}
    \label{eq:H2}
    \mathcal{H} = \sum_{\mathbf p, \alpha} \varepsilon_{\mathbf p}\, c^{\dagger}_{\mathbf p, \alpha} c^{\phantom{;}}_{\mathbf p, \alpha}
    - \frac{g}{V} \!
    \sum_{\mathbf p, \mathbf k, \alpha\beta}\!\! \mathbf B(\mathbf k)\cdot\bm\sigma_{\alpha\beta}\, c^{\dagger}_{\mathbf p, \alpha}
     c^{\phantom{;}}_{\mathbf p + \mathbf k, \beta},\quad
\end{eqnarray}
where $\varepsilon_{\mathbf p}$ is the electron energy measured
with respect to the chemical potential $\mu$, $g \propto J_H$
is the coupling constant, and the correlation for the
magnetization field is given by Eq.~(\ref{eq:B-corr}). For simplicity, here we consider
a parabolic dispersion $\varepsilon_{\mathbf p} = p^2/2m - \mu$, where $m$ is the effective electron mass.
Note that the Hamiltonian is weakly time-dependent owing to the slowly varying magnetic field.
We first calculate the lifetime of electrons close to the Fermi surface. To this end, we consider the leading second-order terms of the
electronic self-energy corresponding to the diagram shown in
Fig.~\ref{fig:self-energy}. Here the solid line denotes the free
electron propagator $G_0(\mathbf p,\omega) = (\omega - \varepsilon_{\mathbf p} \pm i 0)^{-1}$ with $\pm$ sign
corresponding to advanced/retarded Green's function, respectively,
and the wavy line represents the $\mathbf B$-field correlator.
Separating the correlator into the transverse and longitudinal components, we obtain
a self-energy $\Sigma_{a, \,\alpha\beta} = \delta_{\alpha\beta} \Sigma_a$ whose imaginary part is given by
\begin{eqnarray}
    \label{eq:Im_Sigma}
    & &{\rm Im}\Sigma_a(\mathbf p, \omega) = \frac{g^2}{2\pi^2 v K}\sum_{s=\pm1}\int_0^{\Lambda} dk
    \,\,k\,\mathcal{D}_a(k) \nonumber \\
    & & \qquad \times \tan^{-1}\left\{\tau_a[v k - s(\omega - \varepsilon_p - k^2/2m)]\right\},
\end{eqnarray}
Here $a = T$, $L$ labels the transverse and longitudinal components of the
correlator, $v = p/m$ is the electron velocity, $\Lambda = \ell_c^{-1}$ is an ultraviolet cutoff, $\mathcal{D}_T = 2$ 
and $\mathcal{D}_L = \kappa^2/(k^2 + \kappa^2)$. 
The magnetic relaxation times are $\tau_T = \tau_m$ and $\tau_L = 1/(\tau_m^{-1} + D |\mathbf k|^2)$.
Since the correlation~(\ref{eq:B-corr}) is valid only
for coarse-grained magnetizations, the length scale $\ell_c$ should be over several lattice constants.
A natural choice for the cutoff is the Fermi wavevector $p_F$. We note that our results do not depend on the exact 
value of the cutoff, as we show below.

\begin{figure}
\includegraphics[width=0.97\columnwidth]{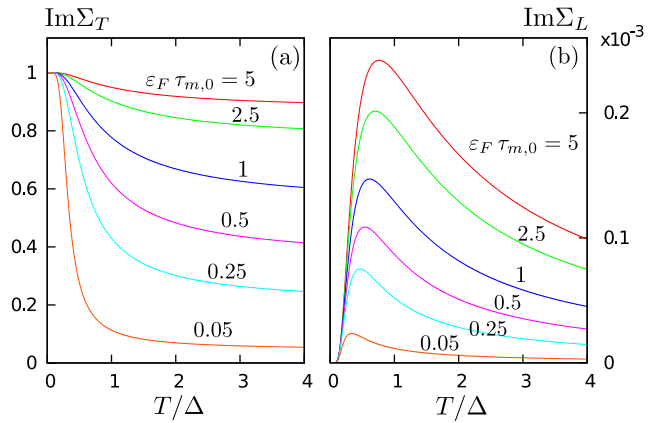}
\caption{(Color online) (a) Transverse and (b) longitudinal components of the
electron scattering rate $\tau_e^{-1}(T) = 2\,{\rm Im}\Sigma$ as a function of temperatue. Both are normalized
with respect to ${\rm Im}\Sigma_T(T = 0) = g^2\Lambda^2/2\pi v_F K$. We have used $\tau_{m,0} D\,p_F^2 = 0.1$
and a Debye screening length $\ell_D(T = \Delta) = 1.65\times 10^3\,\ell_c$ in the calculations.
\label{fig:sigma}}
\end{figure}

We now consider the decay rate $\tau^{-1}_e = 2\,{\rm Im}\Sigma$ of electrons on the Fermi surface $\omega = \varepsilon_p = 0$.
Fig.~\ref{fig:sigma} shows the transverse and longitudinal components of ${\rm Im}\Sigma$ as a function of temperature. In
particular, the integral~(\ref{eq:Im_Sigma}) for the transverse part depends only on two dimensionless parameters $\Lambda/p_F$
and $\varepsilon_F\, \tau_{m,0}$, where $\varepsilon_{F}$ is the Fermi energy.  
The longitudinal part, on the other hand, depends also on the monopole screening length $\ell_D$.
However, since at low temperature $\ell_{D}\gg\ell_{c}$ and $\mathcal{D}_{L}\ll\mathcal{D}_{T}$
for most of the integration region, the longitudinal contribution to the decay rate ${\rm Im}\Sigma$ 
is negligible compared with the transverse part.

Interestingly, the transverse part ${\rm Im}\Sigma_T$ {\em decreases} with increasing $T$.
This is in stark contrast to the behavior of the scattering rate due to other common processes in metals, 
such as scattering by phonons, electrons, or spin fluctuations.
The anomalous temperature dependence of ${\rm Im}\Sigma$ can be understood as follows.
At $T = 0$, a constant relaxation time results from scattering of electrons by a dipolar-correlated static random field.
At finite temperatures, thermally excited monopoles reduce the spatial spin correlation and suppress the electron
scattering rate. In particular, expanding ${\rm Im}\Sigma_T(\mathbf p_F, \omega = 0)$ at low temperatures gives
\begin{equation}
	{\rm Im}\Sigma_{T}=\frac{g^{2}\Lambda^{2}}{2\pi v_{F}K}\left(1-\frac{4p_{F}^{2}}{\pi\Lambda^{2}}\,
	\frac{e^{-\Delta/T}}{\varepsilon_{F}\tau_{m,0}}\tanh^{-1}\!\frac{\Lambda}{2p_{F}}+\cdots\right),
\end{equation}
that decreases exponentially at small $T$.
It should also be noted that at high temperatures $T \gg \Delta$, the correlator~(\ref{eq:B-corr}) of the
coarse-grained magnetization is no longer valid due to proliferation of monopoles. Instead, a power-law relaxation 
rate $\tau^{-1}_e \propto T^n$ results from electron-phonon or electron-magnon scatterings.
The competition of these two scattering mechanisms thus gives rise to a resistivity minimum.

We now turn to the calculation of the resistivity in the Coulomb phase, using the so-called memory function approach~\cite{gotze72}.
In this method the conductivity is expressed as $\sigma(z,T)=(i\omega_{p}^{2}/4\pi)[z+M(z,T)]^{-1}$,
where $\omega_{p}^{2}=4\pi e^{2}n_{e}/m$ is the electronic plasma frequency, $n_{e}$ and $m$ are the electron density and mass, respectively,
and the \emph{memory function} is approximated as $M(z)=(m/n_{e}z)\left[\phi(z)-\phi(0)\right]$ with 
\begin{eqnarray}
    \label{eq:phi}
    \phi(z) = -i \int_0^{\infty} dt\, e^{i z t} \left\langle \big[[\mathcal{H}, j]_t, [\mathcal{H}, j]_{t=0}\big]\right\rangle.
\end{eqnarray}
Here $j = j_\mu = \sum_{\mathbf p} v_{\mu}(\mathbf p) c^{\dagger}_{\mathbf p, \alpha} c^{\;}_{\mathbf p, \alpha}$
is the current operator ($\mu = x, y, z$), $v_{\mu}(\mathbf p) = \partial \varepsilon_{\mathbf p}/\partial p_\mu = p_\mu/m$,
the Hamiltonian $\mathcal{H}$ is given by Eq.~(\ref{eq:H2}),
and $\langle \cdots\rangle$ denotes average over the electronic ground state and degenerate spin configurations
with correlations specified by Eq.~(\ref{eq:B-corr}).

\begin{figure}[b]
\includegraphics[width=0.82\columnwidth]{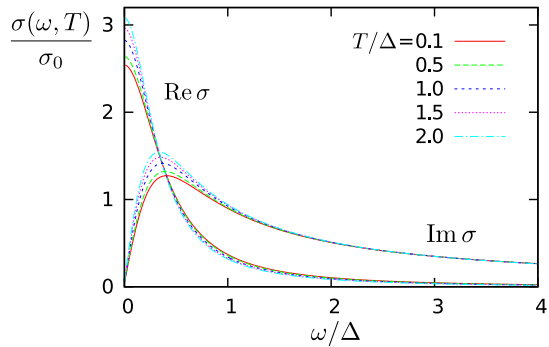}
\caption{(Color online) Dynamical conductivity $\sigma(\omega,T)$ as a function of frequency $\omega$
at various temperatures. The conductivity is measured with respect to $\sigma_0 = \omega_p^2/4\pi\Delta$.
\label{fig:conductivity}}
\end{figure}

\begin{figure}
\includegraphics[width=0.95\columnwidth]{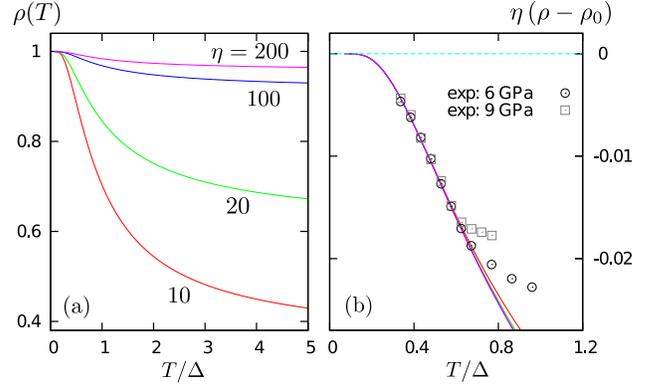}
\caption{(Color online) (a) Resistivity $\rho(T)$ (in units of $\rho_{0}$) as a function of temperature $T$ for varying dimensionless 
parameter $\eta = \varepsilon_F\,\tau_{m,0}$. The Fermi energy is fixed at $\varepsilon_F = 100 \Delta$ in the calculation.
(b) The rescaled resistivity $\eta (\rho - \rho_0)$ collapses on a universal curve at low temperatures.
Also shown is the rescaled experimental data from the measurements on Nd$_2$Ir$_2$O$_7$~\cite{sakata11}.
\label{fig:resistivity}}
\end{figure}

Since, as discussed above, the scattering is dominated by the transverse part of the spin correlation,
we thus focus on its contribution to the conductivity in the following discussion. A straightforward calculation gives
\begin{eqnarray}
    \label{eq:M}
    & & M(z) = \frac{4 g^2}{3 m n_e K}\int \frac{d^3\mathbf p}{(2\pi)^3}\frac{d^3\mathbf k}{(2\pi)^3}
    \,|\mathbf k|^2  \\
    & & \quad \times \frac{f(\varepsilon_{\mathbf p + \mathbf k}) - f(\varepsilon_{\mathbf p})}
    {\left(z + \varepsilon_{\mathbf p} - \varepsilon_{\mathbf p + \mathbf k} + i/\tau_m\right)
    \left(\varepsilon_{\mathbf p + \mathbf k} - \varepsilon_{\mathbf p} - i/\tau_m\right)}, \nonumber
\end{eqnarray}
where $f(\varepsilon_{\mathbf k})$ denotes the Fermi function of energy $\varepsilon_{\mathbf k}$.
Fig.~(\ref{fig:conductivity}) shows the resulting dynamical conductivity as a function of frequency $\omega$ for various temperatures.
By expanding $M(z)$ at low temperatures in powers of $\omega$ and $\exp(-\Delta/T)$, we obtain the standard Drude Lorentzian
form~\cite{pines}: $\sigma(\omega)=\mathfrak{a}\,(\omega_{p}^{2}/4\pi)(i\omega+\tau_{{\rm tr}}^{-1})/(\omega^{2}+\tau_{{\rm tr}}^{-2})$,
where $\mathfrak{a}=(1+\partial M'/\partial\omega|_{\omega=0})^{-1}$ and $\tau_{{\rm tr}}=1/\mathfrak{a}M''(0)$ is the transport relaxation
time. The dc resistivity is related to the transport lifetime as $\rho(T)=(4\pi/\mathfrak{a}\,\omega_{p}^{2})\,\tau_{{\rm tr}}^{-1}(T)$.
Fig.~\ref{fig:resistivity}(a) shows the dc resistivity $\rho(T)$ vs temperature for various dimensionless 
parameter $\eta = \varepsilon_{F}\,\tau_{m,0}$. Similar to the electronic relaxation time, the resistivity increases
to a maximum value as $T$ tends to zero. At low temperatures we again find that $\rho(T)$ decreases exponentially with $T$: 
\begin{eqnarray}
\rho(T)=\rho_{0}\left(1-\frac{8}{3\pi}\frac{p_{F}}{\Lambda}\frac{e^{-\Delta/T}}{\varepsilon_{F}\tau_{m,0}}+\cdots\right),\label{eq:rhoT}
\end{eqnarray}
where $\rho_{0}\propto(g^{2}/K)(\rho_{F}^{2}\varepsilon_{F}/n_{e}\omega_{p}^{2})$ is the residual resistivity at zero temperature.

To leading linear approximation, we use the Matthiessen's rule to combine various contributions to the resistivity.
In particular, we consider a temperature dependence: $\rho_{\rm tot}(T) \sim \rho_0 - \mathcal{A}\, e^{-\Delta/T} + \mathcal{B}\, T^{n}$,
where $\mathcal{A}$ is the coefficient of the exponential term in (\ref{eq:rhoT}) and $\mathcal{B}>0$. For example,
the exponent $n = 2$ for scattering due to electron-electron interaction in a Fermi liquid.
We find that the resistivity has a minimum at
\begin{equation}
    T_{\rm min} = \frac{\Delta/(n+1)}{\left| W\left(\frac{-\Delta}{n+1}
    \left(\frac{\mathcal{B}\,n}{\mathcal{A}\Delta}\right)^{\frac{1}{n+1}}\right)\right|},
\end{equation}
for $\mathcal{A} > \mathcal{A}_c = n \mathcal{B}/\Delta\left(\Delta/(n+1) e\right)^{n+1}$.
Here $W(x)$ denotes the Lambert $W$-function. Since $-1 < W(-x) < 0$ for $x >0$, the resistivity
minimum thus occurs at temperature scales associated with the magnetic monopole excitation energy,
consistent with the numerical calculation in Ref.~\cite{udagawa12}.

Experimentally, the resistivity upturn observed in Nd$_2$Ir$_2$O$_7$ under pressure
agrees well with our calculations. Eq.~(\ref{eq:rhoT}) shows that the low-$T$ resistivity curves $\eta(\rho - \rho_0)$
rescaled by the dimensionless parameter $\eta=\varepsilon_{F}\,\tau_{m,0}$
collapse on a universal curve independent of the magnetic relaxation time. By plotting the rescaled experimental 
resistance $R_{\,\rm exp} - R_0$ as a function of $T/\Delta$, the data points indeed fall on the theoretical curve at low 
temperatures [Fig.~\ref{fig:resistivity}(b)]. From this analysis, we obtain a monopole activation energy $\Delta \approx 5.2$~K, independent 
of applied pressures, and a residual resistance $R_0 \approx 0.87\,\Omega$ and $0.52\,\Omega$ for samples under pressure 6 and 9~GPa, 
respectively. The ratio of the dimensionless parameters $\eta(P)$ at the two pressure values $P$ is $\eta(6\,{\rm GPa})/\eta(9\,{\rm GPa})=3$, 
i.e. the prefactor of the magnetic relaxation rate $1/\tau_{m,0}$ increases by a factor of three as the pressure is increased from 6
to 9 GPa. This is not unexpected, since the spin-spin separation shrinks under pressure and, consequently, the exchange coupling increases. 
We also note that, while the resistivity minimum observed in the other compound Pr$_{2}$Ir$_{2}$O$_{7}$ is compatible with our model prediction,
the low temperature data seems to suggest a $\ln T$ behavior~\cite{nakatsuji06}, implying the effect of partial Kondo-screening.

In summary, we have shown that electronic transport in spin ice exhibits a resistivity minimum
that does not originate from Kondo screening. Instead, it comes from the scattering of electrons by
a random magnetization field with long-range correlation, which is controlled by thermally excited topological defects,
or emergent magnetic monopoles. Using the memory function approach, we have shown that the dc resistivity decreases 
with increasing temperatures, in agreement with experiment and with numerical simulations of the Kondo-lattice model. 
Our theory opens a new route to explore unusual transport phenomena in metallic geometrically frustrated magnets.
Questions that remain open, and deserve further study, are how the magnetic relaxation rate is affected by quantum 
spin dynamics and the feedback effect of the itinerant electrons on the local moments correlations.

{\em Acknowledgement.} We gratefully acknowledge insightful discussions with C.~D.~Batista, A.~V.~Chubukov, A.~Kamenev, I.~Martin, R.~Moessner,
Y.~Motome, and M.~Udagawa. G.W.C. thanks the support of the LANL Oppenheimer Fellowship.

\end{document}